\documentclass[journal,10pt]{IEEEtran}

\usepackage[T1]{fontenc}% optional T1 font encoding
\usepackage{cite,amssymb,color,textcomp,tabularx}
\usepackage{graphicx,epstopdf}
\usepackage{amsmath}
\usepackage[cmintegrals]{newtxmath}
\usepackage{bm}
\usepackage{algorithm,algorithmic}
\usepackage[caption=false,font=footnotesize]{subfig}
\usepackage{stfloats}
\usepackage{multirow}%tabular cells with multiple rows
\usepackage{enumerate,extarrows}
\usepackage{dsfont}% 1(.) indicator function
\usepackage[mathcal]{euscript}

\newtheorem{theorem}{\underline{Theorem}}%[section]
\def \st {\mathrm{s.t.}}
\def \L {\mathcal{L}}
\def \B {\mathcal{B}}
\def \P {\mathbb{P}}
\def \E {\mathbb{E}}

\def \maxp {\mathop{\mathrm{maximize}}}

\newcommand{\mr}[1]{\mathrm{#1}}
\newcommand{\mc}[1]{\mathcal{#1}}

\begin{document}
\title{\large\bfseries Delivery-Secrecy Tradeoff for Cache-Enabled Stochastic Networks:\\Content Placement Optimization}
\author{Qian~Yang,~Hui-Ming~Wang,~and~Tong-Xing~Zheng%
	\thanks{Copyright (c) 2015 IEEE. Personal use of this material is permitted. However, permission to use this material for any other purposes must be obtained from the IEEE by sending a request to pubs-permissions@ieee.org.}%
	\thanks{The authors are with the Ministry of Education Key Lab for Intelligent Networks and Network Security, Xi'an Jiaotong University, Xi'an 710049, China (e-mail: yangq36@gmail.com; xjbswhm@gmail.com; zhengtx@mail.xjtu.edu.cn).}%
	\vspace{-15pt}
}
\begin{@twocolumnfalse}
\maketitle
\end{@twocolumnfalse}
\begin{abstract}
	Wireless caching has been widely recognized as a promising technique for efficient content delivery. % nowadays. 
	In this paper, by taking different file secrecy levels into consideration, physical-layer security oriented content placement is optimized in a stochastic cache-enabled cellular network. 
	We propose an analytical framework to investigate the nontrivial file delivery-secrecy tradeoff.
	Specifically, we first derive the closed-form expressions for the file hit and secrecy probabilities. The global optimal probabilistic content placement policy is then analytically derived in terms of hit probability maximization under file secrecy constraints. 
	Numerical results are demonstrated to verify our analytical findings and show that the targeted file secrecy levels are crucial in balancing the file delivery-secrecy tradeoff.
\end{abstract}


%\IEEEpeerreviewmaketitle
%\newpage

\section{Introduction}\label{sec_intro}
%\IEEEPARstart{T}{his} 
By distributing content across multiple network nodes, caching is an effective way to shift the huge traffic from peak to off-peak hours in cellular networks with the increasing popularity of multimedia streaming and sharing\cite{Maddah-Ali2014,Bastug2014}.
%caching is a promising way to alleviate the increasingly pronounced traffic congestion issue in the backhaul network\cite{}. 
Many efforts have been devoted to investigating efficient content placement in a variety of cache-enabled networks so as to reduce delivery latency \cite{Blaszczyszyn2015,Chae2016,Chae2017,Cui2016,Wen2017}. % and to meet the quality of service requirements.
In \cite{Blaszczyszyn2015}, an optimal randomized content placement policy is proposed to maximize the cache hit probability in a wireless cellular network.
The optimal caching probabilities are derived in \cite{Chae2016} by further considering the channel selection diversity and network interference.
In \cite{Chae2017}, the tradeoff between the content diversity and cooperative gains is investigated.
%In \cite{Chae2017}, a probabilistic content placement scheme is proposed to optimally balance the tradeoff between the content diversity and cooperative gains.
The research on content placement is also extended to multicasting\cite{Cui2016} and heterogeneous\cite{Wen2017} networks.
%The content placement probability is also optimized in cache-enabled multicasting networks\cite{Cui2016} and heterogeneous cellular networks\cite{Wen2017}.

Apart from the aforementioned focus on the efficient content delivery, the content security issues like being eavesdropped by non-paying subscribers or malicious attackers have drawn increasing attention. % due to the broadcast nature of wireless channels.
However, the study on these security issues in cache-enabled networks is still in its infancy.
Most of the existing work as in \cite{Sengupta2015,Gabry2016} aims to improve the network security based on the coded caching scheme proposed in \cite{Maddah-Ali2014}, while leaving the physical-layer channel dynamics unexplored in the context of physical-layer security.
Note that physical-layer security, as an alternative or complement to cryptography, has drawn much attention in ensuring the security of wireless communications since it can guarantee unbreakable (perfect) secrecy\cite{Bloch2011}.
Only in a very recent contribution \cite{Xiang2018}, a cooperative caching scheme based on the physical-layer security is proposed to safeguard video streaming in a backhaul-limited network.
However, only a fixed network topology is considered and the global channel state information is assumed in \cite{Xiang2018}.
Besides, the cached files can have \emph{different secrecy levels} such as the different access permission types for paid video-streaming services, which is largely ignored by most of the existing literature.
To the best of our knowledge, the physical-layer security issue in the cache-enabled stochastic networks with dynamic content-centric user association and files of different secrecy levels has not been studied before. 
%Most importantly, the \emph{file secrecy feature} is generally ignored by most of the existing work, and how to find the optimal caching policy in a stochastic network with files of different secrecy levels is challenging and remains unknown.
Moreover, there exists a nontrivial tradeoff in dealing with \emph{two basic file properties, namely file popularity and file secrecy}, in terms of maximizing the hit probability against eavesdropping.
For instance, it can be unfavorable to cache popular files across many cache-enabled nodes when these files are subject to a high secrecy level, otherwise the file security would be easily compromised.
Therefore, how to find the optimal caching policy under this scenario is of great importance yet remains unknown. %This motivate our work.

%In this paper, we propose an analytical framework to tackle the file delivery-secrecy tradeoff and derive the optimal content placement policy in a stochastic cellular network with files of different secrecy levels.
%In particular, we first introduce guard zones for secrecy enhancement, and then the file hit and secrecy probabilities are obtained and approximated in the closed form, respectively. 
%Finally, the optimal content placement probabilities are analytically obtained by maximizing the hit probability under file secrecy constraints. 
In this paper, we propose an analytical framework to tackle the file delivery-secrecy tradeoff and derive the optimal content placement policy in a stochastic cellular network with files of different secrecy levels for the first time.
Different from \cite{Blaszczyszyn2015,Chae2016,Chae2017,Cui2016,Wen2017}, our work takes the property of different file secrecy levels into account and uncovers the intrinsic delivery-secrecy tradeoff, which further yields the results fundamentally different from the conventional ones.
Many new insights and the impacts of various network parameters on the nontrivial delivery-secrecy tradeoff are unveiled in our paper.

\section{Network Model}\label{sec_net_model}
%\begin{figure}[t]
%\centering
%\includegraphics[width=5in]{topo}
%\caption{The stochastic B-WPC ad hoc network consisting of multiple RFID reader-tag pairs.
%	}
%\label{fig_topo}
%\end{figure}
\subsection{Network Topology}\label{subsec_net_topo}
We consider a stochastic cache-enabled cellular network, where the locations of base stations (BSs) are modeled as a two-dimensional homogeneous Poisson point process (PPP) $ \Phi $ with density $\lambda$. 
The user density in the network is assumed to be much larger than the BS density such that all the BSs are active as in \cite{Chae2016,Wen2017}. 
%The number of the users in the network is assumed to be much larger than that of BSs. Thus, each BS is active and serves one user in the network.
Additionally, there are multiple passive eavesdroppers (e.g., non-paying subscribers) aiming to wiretap different kinds of legitimate transmission (e.g., paid video-streaming services corresponding to different membership grades) with different secrecy levels. 
The locations of all the eavesdroppers are represented by another homogeneous PPP $\Phi_e$ with density $\lambda_e$.
Each network node has a single antenna.

To alleviate backhaul pressure for file requests from the network users, each BS is equipped with a local cache with a predetermined storage size $C$.
The size of each file (or file fragment) is normalized to the unit size, and at most $C$ files (or fragments) can be cached at each BS.
%accordingly.

\subsection{File Properties}\label{subsec_file}
The total number of different confidential files in the network is $ F $ ($F>C$), and the files are denoted by the collection $ \mc{F}=\{1,\ldots,F\} $.
Each file $f\in \mc{F}$ has two kinds of properties as follows. For one thing, as commonly adopted in \cite{Bastug2014,Blaszczyszyn2015,Chae2016,Chae2017,Cui2016,Wen2017}, each user requests all the $F$ files with a given popularity distribution. % which is usually fixed for a long period and can be perfectly known. 
For instance, under the Zipf distribution the probability for each user to request file $f\in \mc{F}$ is given as 
\begin{equation}
q_f=\frac{1/f^\beta}{\sum_{i=1}^F 1/i^\beta},\quad \forall f\in \mc{F},
\end{equation}
where $\beta\geq 0$ denotes the skewness of the popularity distribution. We will use this popularity distribution hereinafter.
%In particular, larger $\beta$ implies a more diverse distribution, and the distribution becomes uniform when $\beta=0$.
%Note that the file with lower index has higher popularity under the Zipf distribution, i.e., $q_i\geq q_j$ if $i<j$, and the discussion in this paper can be easily extended to the scenarios with any other discrete popularity distributions.

%For another thing, each file can have different secrecy levels as another file property. 
For another thing, each file can have different secrecy levels as another file property, which is not considered in \cite{Blaszczyszyn2015,Chae2016,Chae2017,Cui2016,Wen2017}.
This property is of great importance in a lot of commercial and military applications. %, apart from the aforementioned popularity property.
For instance, in video streaming services, some video files can be of high secrecy levels and available to only certain users with qualified membership, while other files may be not. % and can be even freely accessed by the public.
In this paper, we propose to use different secrecy constraints to account for different file secrecy levels. 
The secrecy constraint for file $i$ is characterized by
\begin{equation} \label{secrecy_cons}
\mc{P}_{si}\geq \epsilon_i,~\forall i\in\mc{F},
\end{equation}
where $\mc{P}_{si}$ denotes the secrecy probability (will be defined in Section \ref{subsec_secrecy_prob}) for the transmission of file $i$, and $\epsilon_i$ is the corresponding secrecy level. A high secrecy level is demanded by imposing a large $\epsilon_i$.

\subsection{Content Delivery and Secrecy Schemes}\label{subsec_scheme}
As in \cite{Blaszczyszyn2015,Chae2016,Chae2017,Cui2016,Wen2017}, we adopt a \emph{probabilistic content placement} policy for caching. Specifically, each BS independently caches file $i$ with probability $p_i$ for $i\in \mc{F}$.
The caching probabilities $\{p_i\}_{i=1}^F$ are subject to the cache storage constraint given by
\begin{equation}\label{storage_cons}
\sum_{i=1}^{F}p_i \leq C.
\end{equation}
As illustrated in \cite{Blaszczyszyn2015}, once given \eqref{storage_cons} there always exists a specific caching scheme %(e.g., as described in Fig. 1 of \cite{Blaszczyszyn2015}) 
satisfying the per-BS cache storage constraint.
In this regard, how to optimize $\{p_i\}_{i=1}^F$ under file secrecy constraints is the main focus of this paper.

Once the caching process is fulfilled during off-peak hours, the locations of the BSs with file $i$ available can be represented as a PPP $\Phi_i$ with density $\lambda_i\triangleq p_i\lambda$ for $i\in \mc{F}$ according to the thinning theory of PPP\cite{Chiu2013}. 
The BSs without caching file $i$ are thereby characterized by another independent PPP $\Phi_i^c$ with density $\lambda_i^c\triangleq(1-p_i)\lambda$.
All the channels from the BSs and to users are assumed to be quasi-static and undergo small-scale Rayleigh fading combined with large-scale path loss.
We consider a content-centric association policy as in \cite{Cui2016,Wen2017}. In particular, any user requesting file $i$ will be associated to the nearest BS in $\Phi_i$. 
%When multiple files are available at a BS, multiple users may associate to it, and the BS transmits each on-demand file in a time-division-multiple-access (TDMA) way for each associated users.
When multiple files are available at a BS, multiple users may associate to it, and the BS transmits each on-demand file in a round-robin scheduling manner.

In this paper, we consider the scenario where the eavesdroppers are the internal users of the network but have no access to the confidential files.
A non-colluding and eavesdropper-centric wiretapping scenario is considered.
For a typical eavesdropper, it tries to individually decode each confidential file transmitted from the closest BS with the file available. %TODO justify
To safeguard communication security, we assume that a secrecy guard zone is set around each BS as in \cite{Zhou2011}.
The guard zone is modeled as a disk with radius $D$ centered at each BS, where as commonly assumed each BS can individually detect the existence of eavesdroppers through their location measurements or a special power-aware medium access control (MAC) protocol \cite{Hasan2007} since the eavesdroppers are internal users. 
%The guard zone is modeled as a disk with radius $D$ centered at each BS, where each BS can individually detect the existence of eavesdroppers. 
Once the eavesdropper is detected, the BS transmits artificial noise (AN) instead of secrecy files with the maximum transmit power for anti-eavesdropping.
%In our paper, the guard zone is a circle with  centered at each BS.

%Performance Metric and Content Placement Problem
\subsection{Problem Formulation for Delivery-Secrecy Tradeoff}\label{subsec_metric}
As pointed out in \cite{Wen2017}, the hit probability, defined as the probability of successful content delivery, is an important performance metric which closely reflects the latency reduction of the packet transmission via a backhaul network. 
Therefore, we aim to maximize the average hit probability $\mc{P}_c$ under file secrecy constraints, which is mathematically represented as
%is chosen as the main performance metric 
\begin{subequations} \label{maxp}
\begin{align}
\maxp_{\{p_i\}_{i=1}^F}~&\mc{P}_{c}\triangleq \sum_{i=1}^{F}q_i \mc{P}_{ci} \label{Pc0}\\
\st~&\eqref{secrecy_cons},~\eqref{storage_cons},~0\leq p_i\leq 1,~\forall i\in\mc{F},
\end{align}
\end{subequations}
where $\mc{P}_{ci}$ is the conditional hit probability given that file $i$ is requested.
Note that problem \eqref{maxp} captures a fundamental delivery-secrecy tradeoff in the cache-enabled network with files of different secrecy levels.  
Intuitively, a large caching probability is beneficial for improving the hit probability, while the security of the file can be more easily compromised by caching across a large number of BSs.

\section{Hit Probability and File Secrecy Probability} \label{sec_prob}
In this section, the file hit and secrecy probabilities are respectively investigated. 
Since the receiver noise is far smaller than the experienced interference in the network, we focus on the network performance in the interference-limited regime.

\subsection{File Hit Probability} \label{subsec_hit_prob}
To calculate the averaged hit probability, we first focus on the hit probability $\mc{P}_{ci}$ conditioned on that file $i$ is requested by a typical user located at the origin based on Slivnyak's theorem \cite{Chiu2013}.
According to the secrecy scheme of guard zones introduced in Section \ref{subsec_scheme}, the set of actual file transmitters for file $ i $ is approximated by a thinned PPP $ \Phi_{ai} $ with density 
\begin{equation}\label{l_ai}
\lambda_{ai}=\lambda_i \exp\left(-\lambda_e \pi D^2\right),
\end{equation}
where the term $ \exp(-\lambda_e \pi D^2) $ denotes the void probability that no eavesdropper is located in the guard zone as in \cite{Zhou2011}.
Similarly, the counterpart of $ \Phi_{ai} $ is denoted by another independent PPP $ \Phi_{\bar{a}i} $ with density $ \lambda_{\bar{a}i}=\lambda_i \left(1-\exp\left(-\lambda_e \pi D^2\right)\right) $.
Note that $ \Phi=\Phi_{ai} \cup \Phi_{\bar{a}i} \cup \Phi_i^c$.

Based on \eqref{l_ai}, the associated BS for the typical user requesting file $ i $ is denoted by $ X_b=\arg \max_{X\in \Phi_{ai}} P \|X\|^{-\alpha} $, where $ P $ is the maximum transmit power of BSs and $ \alpha $ is the large-scale path-loss exponent.
The total interference experienced by the typical user requesting file $ i $ is given by
\begin{equation}\label{I_i}
I(i)= \underbrace{\sum_{X \in \Phi_i^c\cup\Phi_{\bar{a}i}}P h_X \|X\|^{-\alpha}}_{\substack{I_0(i),~\text{interference of}\\ \text{the other files and AN}}}+\underbrace{\sum_{X \in \Phi_{ai}\backslash X_b}P h_X \|X\|^{-\alpha}}_{\substack{I_1(i),~\text{interference from}\\ \text{unassociated BSs with file }i}},
\end{equation}
where the random variable $h_X\sim \mathrm{Exp}(1)$ accounts for the small-scale Rayleigh fading of the channel from BS $X$ to the typical user. {Note that the received signals from the unassociated BSs with file $ i $ available may contain the desired information, but they are incoherent due to longer signal arriving delay and thereby treated as interference $ I_1(i) $ in \eqref{I_i} similar to \cite{Chae2016,Cui2016,Wen2017}.}
Given the targeted signal-to-interference ratio (SIR) $ \gamma_u $, the conditional hit probability $\mc{P}_{ci}$ is defined as the probability that the actual SIR is larger than $ \gamma_u $ for $i\in \mc{F}$.
The hit probability $\mc{P}_{c}$ is given in the following theorem.

\begin{theorem}\label{th_hit}
	The file hit probability $\mc{P}_{c}$ is given as
	\begin{equation}\label{Pc}
	\mc{P}_{c}=\sum_{i=1}^{F}q_i \frac{p_i}{\tau_1(\gamma_u) p_i+\tau_2(\gamma_u)},
	\end{equation}
	with $ \tau_1(\gamma)\triangleq 1+\kappa_2(\gamma)-\kappa_1(\gamma) $, $ \tau_2(\gamma)\triangleq\kappa_1(\gamma) \exp(\pi \lambda_e D^2) $,
	 $ \kappa_1(\gamma)\triangleq \delta \gamma^\delta B(1-\delta,\delta) $, $ \kappa_2(\gamma)\triangleq \frac{\delta\gamma}{1-\delta} \,_2F_1(1,1-\delta;2-\delta;-\gamma) $, and $ \delta\triangleq \frac{2}{\alpha} $, where $B(\cdot)$ and $ {}_2F_1(\cdot) $ denote the beta function \cite[eq. (8.380.1)]{Gradshteyn2007} and the Gauss hypergeometric function \cite[eq. (9.14.2)]{Gradshteyn2007}, respectively.
\end{theorem}
\begin{IEEEproof}
	See Appendix \ref{app_cond_hit}.
\end{IEEEproof}

From Theorem \ref{th_hit}, we find that $\mc{P}_{c}$ increases with the caching probability $ \{p_i\}_{i=1}^F $ aligned with intuition.
%, which is not surprising since the conditional hit probability can be improved by caching more the requested file.
Furthermore, $\mc{P}_{c}$ decreases as the density of the eavesdroppers or the area of the guard zones becomes larger. 
The reason behind this is that due to the setting of guard zones the number of actual file transmitters becomes smaller under these situations.

\subsection{File Secrecy Probability} \label{subsec_secrecy_prob}
%To calculate the secrecy probability of file $ i $, it is sufficient to consider the probability of the secrecy compromised by a typical eavesdropper located at the origin based on Slivnyak's theorem \cite{Chiu2013}.
As in \cite{Zhou2011}, the actual transmitters for file $ i $ can be approximated by a homogeneous PPP with density $ \lambda_{ai} $ outside $ \B(o,D) $ from the viewpoint of the typical eavesdropper located at the origin, where $ \B(o,r) $ denotes a disk of radius $ r $ centered at $ o $.
For all the BSs in $ \Phi\cap\B(o,D) $, they transmit AN to combat eavesdropping according to the guard zone setting. 

The targeted BS for the typical eavesdropper to wiretap file $ i $ is denoted by $ X_e=\arg \max_{X\in \Phi_{ai}\backslash \B(o,D)} P \|X\|^{-\alpha} $.
The probability density function (PDF) of the distance $ \|X_e\| $ can be computed using the void probability of PPP as
\begin{equation} \label{f_Xe}
 f_{\|X_e\|}(r)=2\pi \lambda_{ai}r \exp\left(-\pi\lambda_{ai}\left(r^2-D^2\right)\right), ~ \text{for}~r>D.
\end{equation}
The received interference at the typical eavesdropper\footnote{The method developed in our paper can be easily extended to tackle the case where the eavesdroppers have the ability of joint decoding. Under that case, the eavesdroppers only suffer from AN, which actually takes a degenerate form of \eqref{Ie}.} for wiretapping file $ i $ is formulated by
\begin{align}\label{Ie}
I_e(i)=&\sum_{X \in \Phi_i^c\cup\Phi_{\bar{a}i}}P h_X \|X\|^{-\alpha}+\sum_{X \in \Phi_{ai}\backslash \B(o,D) \backslash X_e}P h_X \|X\|^{-\alpha}\notag\\
&\quad+\sum_{X \in \Phi_{ai}\cap \B(o,D)}P h_X \|X\|^{-\alpha}.
\end{align}
The secrecy probability of file $ i $ is thereby defined as 
\begin{equation}\label{def_Psi}
\mc{P}_{si}=\P\left\{\frac{Ph_{X_e}\|X_e\|^{-\alpha}}{I_e(i)}<\gamma_e \right\},
\end{equation}
where $ \gamma_e $ is the SIR threshold for ensuring the file transmission security. 
Note that the definition in \eqref{def_Psi} corresponds to the case where a wiretap code is constructed by setting the rate redundancy $ R_e=\log(1+\gamma_e) $ to achieve perfect secrecy\cite{Bloch2011}.
Similar to the file hit probability, the expression of $ \mc{P}_{si} $ and one of its lower bounds are provided in the following theorem.
\begin{theorem}\label{th_secrecy}
	The secrecy probability $\mc{P}_{si}$ for file $ i $ is given as
	\begin{align}\label{Psi}
	\mc{P}_{si}=&1-\int_{D}^{\infty} \exp\Bigg(-\pi\left((\lambda_i^c+\lambda_{\bar{a}i})\kappa_1(\gamma_e)+\lambda_{ai}\kappa_2(\gamma_e)\right)r^2\notag\\
	&~~ \underbrace{-\pi\lambda_{ai}D^2 \,_2F_1\left(1,\delta;\delta+1;-\frac{D^\alpha}{\gamma_e r^\alpha}\right)}_{\Theta(r)} \Bigg) f_{\|X_e\|}(r) dr.
	\end{align}
	 One of its lower bounds is obtained by replacing the part $ \Theta(r) $ with $ \Theta(D) $ in \eqref{Psi} and calculating the integral as
	\begin{equation}\label{Psi_LB}
	\mc{P}_{si}>\mc{P}_{si}^{L}\triangleq1-\frac{\exp\left(-\pi D^2\kappa_1(\gamma_e)\lambda\right)}{\tau_1(\gamma_e)+\tau_2(\gamma_e)/p_i}.
	\end{equation}
%	(\kappa_1-\kappa_2-{}_2F_1(1,\delta;\delta+1;-1/\gamma_e))e^{-\pi D^2\lambda_e}p_i
\end{theorem}
\begin{IEEEproof}
	See Appendix \ref{app_secrecy}.
\end{IEEEproof}

\begin{figure}[t]
	\centering
	\includegraphics[width=2.4in]{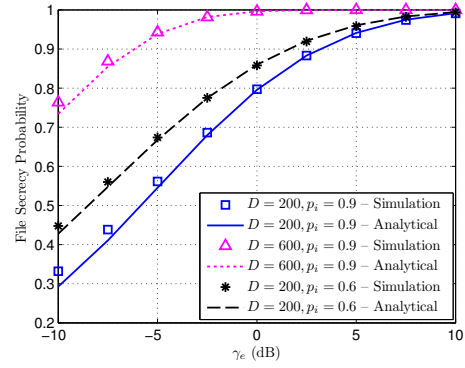}
	\caption{File secrecy probability versus the SIR threshold under different sizes of guard zones and the content placement probabilities of file $ i $, where $ \alpha=3 $, $ \lambda=\frac{1}{800^2\mr{m}^2}$, and $ \lambda_e=\lambda/5 $. }
	\label{fig_gamma_secrecy} 
	\vspace{-10pt}
\end{figure}

Based on the diminishing distribution of $  f_{\|X_e\|}(r) $ in \eqref{f_Xe}, the lower bound given in \eqref{Psi_LB} is expected to be tight especially when $ \lambda_{ai} $ is large.
As validated in Fig. \ref{fig_gamma_secrecy}, the lower bound (labeled as ``Analytical'') coincides well with the simulation results and can be used to approximate $ \mc{P}_{si} $ in constraint \eqref{secrecy_cons}.
From \eqref{Psi_LB} and Fig. \ref{fig_gamma_secrecy}, the secrecy probability for a specific file is a decreasing function of its placement probability as expected, since caching the file across more BSs incurs a higher risk of being eavesdropped.
Moreover, the file secrecy probability increases as the guard zone area becomes larger.

%Hit Probability Maximization
\section{Optimization for Delivery-Secrecy Tradeoff} \label{sec_hit_max}
In this section, we aim to solve problem \eqref{maxp} and find the optimal caching probabilities maximizing the hit probability with files of different secrecy levels.

With $ \mc{P}_{si} $ approximated by $ \mc{P}_{si}^{L} $, problem \eqref{maxp} is changed to
\begin{equation}\label{maxp1}
\maxp_{\{p_i\}_{i=1}^F}~\mc{P}_{c}\qquad 
\st~\eqref{storage_cons},~0\leq p_i\leq \varPsi_i,~\forall i\in\mc{F},
\end{equation}
where $ \varPsi_i\triangleq \min\left\{1,\varPsi_i^\circ\right\} $ with
\begin{equation}\label{varPsi}
\varPsi_i^\circ\triangleq \frac{\tau_2(\gamma_e)(1-\epsilon_i)}{\left[e^{-\pi D^2\kappa_1(\gamma_e)\lambda}-\tau_1(\gamma_e)(1-\epsilon_i)\right]^+}
\end{equation}
accounting for the file secrecy constraints in \eqref{secrecy_cons} and $[x]^+\triangleq \max\{0,x\}$.
It is not hard to see that problem \eqref{maxp1} is convex, and it can be thereby optimally solved by the Lagrange method.
%, since the objective function in \eqref{Pc} is concave and all the constraints are linear.
%Therefore, the problem can be solved optimally by the Lagrange method.

%Specifically, t
The Lagrange function for problem \eqref{maxp1} is given by
\begin{equation}\label{Lagrange}
L(\{p_i\}_{i=1}^F,\nu)=\mc{P}_{c}+\nu\left(C-\sum_{i=1}^{F}p_i\right),
\end{equation}
where $ \nu\geq 0 $ is the dual variable corresponding to constraint \eqref{storage_cons}.
Since the objective function in \eqref{maxp1} is an increasing function of $ \{p_i\}_{i=1}^F $, the optimum is given by $ p_i^\star= \varPsi_i,~\forall i\in\mc{F} $ if $ \sum_{i=1}^{F}\varPsi_i \leq C $. 
Otherwise, constraint \eqref{storage_cons} is active at the optimum, i.e., $ \sum_{i=1}^{F}p_i^\star = C $.
By leveraging the Karush-Kuhn-Tucker (KKT) condition, the optimal placement probabilities under $ \sum_{i=1}^{F}\varPsi_i > C $ are given in the following theorem.

\begin{theorem}\label{th_cache}
	Under $ \sum_{i=1}^{F}\varPsi_i > C $, the optimal placement probabilities $ \{p_i\}_{i=1}^F $ for problem \eqref{maxp1} are given by
	\begin{equation}\label{optimal_cache}
	p_i^\star(\nu^\star)=\begin{cases}
	\varPsi_i, &p_i^\circ\geq \varPsi_i,\\
	p_i^\circ,&0<p_i^\circ<\varPsi_i,\\
	0,&p_i^\circ \leq 0,
	\end{cases}
	\end{equation}
	where
	\begin{equation}\label{pi0}
	p_i^\circ \triangleq \frac{1}{\tau_1(\gamma_u)}\sqrt{\frac{\tau_2(\gamma_u)}{\nu^\star}} \sqrt{q_i}-\frac{\tau_2(\gamma_u)}{\tau_1(\gamma_u)},
	\end{equation}
	and the optimal $ \nu^\star $ satisfies $ \sum_{i=1}^{F}p_i^\star(\nu^\star) =C $.
\end{theorem}
\begin{IEEEproof}
%	See Appendix \ref{app_cache}.
According to the KKT condition, when $ 0<p_i^\star(\nu^\star)<\varPsi_i $ the optimal placement probabilities can be found as $ p_i^\star(\nu^\star)=p_i^\circ $ by setting the first-order derivative of the Lagrange function in \eqref{Lagrange} to zero.
The two boundary results are obtained using the complementary slackness condition\cite{Boyd2004}.
The proof is completed.
\end{IEEEproof}

\begin{figure}[t]
	\centering
	\includegraphics[width=2.2in]{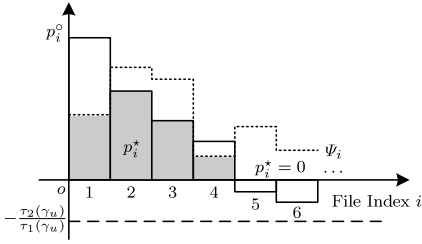}
	\caption{An illustration for the optimal cache placement strategy under different file secrecy levels, where the shadowed parts stand for the optimal placement probability.}
	\label{fig_opt_cache} 
	\vspace{-10pt}
\end{figure}

Since $ p_i^\circ $ in \eqref{pi0} is a deceasing function of $ \nu^\star $, the optimal $ \nu^\star $ can be found by a bisection search.
%From Theorem \ref{th_cache}, we find that the placement probability cannot be set arbitrarily large due to the file secrecy constraint.
%Because of this, the file with higher popularity does not necessarily have a higher placement probability. To clearly see this, 
According to Theorem \ref{th_cache}, the optimal cache placement strategy under different file secrecy levels is illustrated in Fig. \ref{fig_opt_cache}.
In the figure, $ p_i^\circ $ actually reflects the popularity of file $ i $ since it is an increasing function of $ q_i $ from \eqref{pi0}.
%As observed from Fig. \ref{fig_opt_cache}, even though the popularity is diminishing as the file index becomes larger, the same trend does not hold for the optimal placement probability due to the secrecy property of each file captured by $ \{\varPsi_i\}_{i=1}^F $.
Unlike the results in \cite{Blaszczyszyn2015,Chae2016,Chae2017,Cui2016,Wen2017}, even though the popularity is diminishing as the file index becomes larger, the same trend does not hold for the optimal placement probability due to the secrecy property of each file captured by $ \{\varPsi_i\}_{i=1}^F $.

%Since $ p_i^\circ $ in \eqref{pi0} is a deceasing function of $ \nu^\star $, the optimal $ \nu^\star $ can be found by a bisection search.
%From Theorem \ref{th_cache}, we find that the placement probability cannot be set arbitrarily large due to the file secrecy constraint.
%Because of this, the file with higher popularity does not necessarily have a higher placement probability. To clearly see this, the optimal cache placement strategy under different file secrecy levels is illustrated in Fig. \ref{fig_opt_cache}.
%In the figure, $ p_i^\circ $ actually reflects the popularity of file $ i $ since it is an increasing function of $ q_i $ from \eqref{pi0}.
%As observed from Fig. \ref{fig_opt_cache}, even though the popularity is diminishing as the file index becomes larger, the same trend does not hold for the optimal placement probability due to the secrecy property of each file captured by $ \{\varPsi_i\}_{i=1}^F $.

\section{Numerical Results}\label{sec_sim}
Some numerical results are presented in this section to show the superiority of the proposed optimal content placement (OCP) strategy.
% under the scenario with files of different secrecy levels.
The benchmark schemes include caching the most popular contents (MPC) and caching the least classified contents (LCC) under file secrecy constraints.\footnote{Similar to the MPC scheme following the descending order of $q_i$, the LCC scheme follows the ascending order of $\epsilon_i$ during the caching process.}
The parameter settings are as follows, unless otherwise specified: $ \alpha=3 $, $ \lambda=\frac{1}{800^2\mr{m}^2}$, $ \lambda_e=\lambda/5 $, $ C=5 $, $ F=10 $, $ D=200~\mr{m} $, $ \gamma_u=-5 $ dB, $ \gamma_e=-7 $ dB, and the file secrecy levels are independently generated as $ \epsilon_\mr{max} \mathtt{Rand(1)}$ with $\mathtt{Rand(1)}$ denoting a random variable uniformly distributed in $(0,1)$. %, and all the simulation results are obtained by averaging over $ 10,000 $ network realizations. 

% beta, eps_max

\begin{figure}[t]
	\centering
	\includegraphics[width=2.36in]{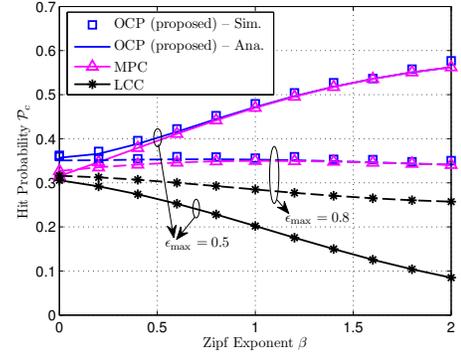}
	\caption{The hit probability versus the Zipf exponent under different content placement schemes.}
	\label{fig_beta_hit}
	\vspace{-10pt}
\end{figure}

In Fig. \ref{fig_beta_hit}, the performance of the proposed OCP scheme is compared with the results obtained by MPC and LCC.
It can be observed from Fig. \ref{fig_beta_hit} that the proposed scheme yields the largest hit probability and the analytical results coincide well with the simulation ones. %, which verifies the correctness of our theoretical derivations. 
Moreover, according to Fig. \ref{fig_beta_hit} the performance of MPC is close to the optimal one when the content popularity is skewed (a larger $\beta$) as expected.
Contrarily, the hit probability obtained by LCC is larger when the content popularity becomes uniform (a smaller $\beta$) due to the absence of the popularity information in carrying out LCC.
Furthermore, from Fig. \ref{fig_beta_hit} the optimal hit probability decreases as $ \epsilon_\mr{max} $ grows, which reflects the fundamental tradeoff between content delivery and content secrecy.
However, the performance obtained by LCC can be improved by imposing more stringent secrecy constraints, since there can be more storage left for caching the other popular files.
It is also interesting to notice that the optimal hit probability is not necessarily an increasing function of $ \beta $ under the secrecy-limited case ($ \epsilon_\mr{max}=0.8 $) from Fig. \ref{fig_beta_hit}. 
%For , the optimal performance becomes a little worse as $ \beta $ gets larger. This is because ensuring a high secrecy level for the most popular file can considerably limit its caching probability, which thereby prevents the total hit probability from becoming larger.

\begin{figure}[t]
	\centering
	\includegraphics[width=2.4in]{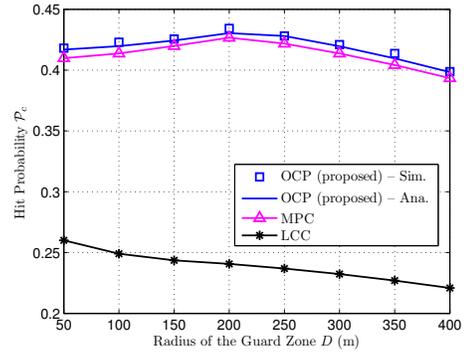}
	\caption{The hit probability versus the radius of guard zones under different content placement schemes with $ \epsilon_\mr{max}=0.5 $ and $ \beta=0.7 $.}
	\label{fig_D_hit}
	\vspace{-10pt}
\end{figure}

The impact of the guard zone radius $ D $ on the hit probability is shown in Fig. \ref{fig_D_hit}.
It is observed that there exists an optimal $ D $ for maximizing the hit probability.
Actually, this parameter captures the tradeoff between file hit and secrecy probabilities, since a large $ D $ reduces the density of actual file transmitters and thereby the file hit probability while it strengthens file secrecy.
However, for the LCC scheme, the hit probability monotonically decreases with an increase in $ D $ from Fig. \ref{fig_D_hit}. The reason behind is that the file secrecy levels for the cached files are low. Therefore, the effect of $ D $ on file secrecy is less pronounced compared with content delivery.

\section{Conclusion} \label{sec_conclusion}
In this paper, we have studied the security issues in a cache-enabled stochastic network. % with files of different secrecy levels.
In particular, an analytical framework has been proposed to tackle the file delivery-secrecy tradeoff. % in a stochastic network for the first time.
Based on the closed-form expressions of the file hit and secrecy probabilities,  we have analytically obtained the optimal content placement policy in terms of hit probability maximization. 
Numerical results show the superiority of the proposed scheme and the impacts of file secrecy levels and the guard zone size on the file delivery-secrecy tradeoff.
\begin{appendices}
\section{Proof of Theorem \ref{th_hit}}\label{app_cond_hit}
Based on \eqref{I_i}, the conditional hit probability $\mc{P}_{ci}$ for the typical user requesting file $ i $ is calculated as
\begin{align}\label{Pci0}
\mc{P}_{ci}
%&=\P\left\{\frac{P h_{X_b} \|X_b\|^{-\alpha}}{I_0(i)+I_1(i)}>\gamma_u\right\}\notag\\
&=\E_{\Phi}\left\{\exp\left(-\frac{\gamma_u \|X_b\|^{\alpha}}{P}\left(I_0(i)+I_1(i)\right)\right)\right\}\notag\\
&=\int_{0}^{\infty} \L_{I_0}\left(\frac{\gamma_u }{P}r^\alpha \right) \L_{I_1}\left(\frac{\gamma_u }{P}r^\alpha \right) f_{\|X_b\|}(r) dr,
\end{align}
where $ f_{\|X_b\|}(r)=2\pi \lambda_{ai}r \exp\left(-\pi\lambda_{ai}r^2\right) $ is the PDF of $ \|X_b\| $.
%the distance between the typical user and its associated closest BS in $ \Phi_{ai} $.

The Laplace transform of $ I_0(i) $ is computed as
\begin{align}\label{L_I0}
\L_{I_0}(s)
&\overset{(a)}{=}\E_{\Phi_i^c,\Phi_{\bar{a}i}}\left\{\prod_{X\in \Phi_i^c\cup\Phi_{\bar{a}i}} \E_{h_X} \left[\exp\left(-P h_X \|X\|^{-\alpha}s \right) \right] \right\}\notag\\
&\overset{(b)}{=}\exp\left(-2\pi(\lambda_i^c+\lambda_{\bar{a}i}) \int_{0}^{\infty}\left(1-\frac{1}{1+sP r^{-\alpha}}\right)rdr \right)\notag\\
&\overset{(c)}{=}\exp\left(-\delta\pi(\lambda_i^c+\lambda_{\bar{a}i}) (sP)^\delta B(1-\delta,\delta)  \right),
\end{align}
where $ (a) $ follows from the independence of different small-scale fading terms, $ (b) $ holds due to the use of the probability generating functional lemma (PGFL) over PPP \cite{Chiu2013}, and $ (c) $ is derived by using the variable change $ x=sP r^{-\alpha} $ and \cite[eq. (3.194.3)]{Gradshteyn2007}.
Similarly, the Laplace transform of $ I_1(i) $ conditioned on $ \|X_b\|=r $ is computed as
\begin{align}\label{L_I1}
%&\L_{I_1}\left(s\big|\|X_b\|=r\right)\notag\\
&\L_{I_1}(s,r)\notag\\
%&\overset{}{=}\E_{\Phi_{ai}\backslash X_b}\left\{\prod_{X \in \Phi_{ai}\backslash X_b} \E_{h_X} \left[\exp\left(-P h_X \|X\|^{-\alpha}s \right) \right] \right\}\notag\\
&\overset{}{=}\exp\left(-2\pi\lambda_{ai} \int_{r}^{\infty}\left(1-\frac{1}{1+sP z^{-\alpha}}\right)zdz \right)\notag\\
&\overset{}{=}\exp\left(-\delta\pi\lambda_{ai} sP\frac{r^{2-\alpha}}{1-\delta} \,_2F_1(1,1-\delta;2-\delta;-sPr^{-\alpha})  \right).
\end{align}

Substituting \eqref{L_I0} and \eqref{L_I1} into \eqref{Pci0} and calculating the integral, we obtain $ \mc{P}_{ci}=\frac{p_i}{\tau_1(\gamma_u) p_i+\tau_2(\gamma_u)} $. 
The hit probability is thereby given in \eqref{Pc} with the aid of its definition in \eqref{Pc0}.
%The proof is completed.

\section{Proof of Theorem \ref{th_secrecy}}\label{app_secrecy}
Based on \eqref{def_Psi}, the secrecy probability for file $ i $ is given by
\begin{equation}\label{Psi2}
\mc{P}_{si}
%&=1-\E_{\Phi}\left\{\exp\left(-\frac{\gamma_e \|X_e\|^{\alpha}}{P}I_e(i)\right)\right\}\notag\\
=1-\int_{D}^{\infty} \L_{I_e}\left(\frac{\gamma_e }{P}r^\alpha \right) f_{\|X_e\|}(r) dr,
\end{equation}
where $ \L_{I_e}(s) $ is the Laplace transform of $ I_e(i) $ condition on $ \|X_e\|=r $.
From \eqref{Ie}, $ \L_{I_e}(s)=\L_{I_0}(s)\cdot\L_{I_1}(s,r)\cdot\L_{I_2}(s) $ with
\begin{align}
\L_{I_2}(s)
&\triangleq \E \, \Bigg\{\prod_{X \in \Phi_{ai}\cap \B(o,D)} \E_{h_X} \left[\exp\left(-P h_X \|X\|^{-\alpha}s \right) \right] \Bigg\}\notag\\
&=\exp\Bigg( -2\pi\lambda_{ai}\int_{0}^{D}\left(1-\frac{1}{1+sP r^{-\alpha}}\right) rdr \Bigg) \notag\\
&\overset{(d)}{=}\exp\left(-\pi\lambda_{ai}D^2 \,_2F_1\left(1,\delta;\delta+1;-\frac{1}{sPD^{-\alpha}}\right)\right),
\end{align}
where $ (d) $ is derived by using the variable change $ x=sP r^{-\alpha} $ and \cite[eq. (3.194.2)]{Gradshteyn2007}. Therefore, with $ \L_{I_e}(s) $ and \eqref{Psi2}, $ \mc{P}_{si} $ takes the form of \eqref{Psi}, of which the closed form is difficult to obtain.
However, since the part $ \Theta(r) $ in \eqref{Psi} is a decreasing function of $ r $, a lower bound of $ \mc{P}_{si} $ is thereby given in \eqref{Psi_LB}.
%The proof is completed.

\end{appendices}

\linespread{0.95} 
%\bibliographystyle{IEEEtran}
%\bibliography{IEEEabrv,my}


\end{document}